\documentclass[preprint,prd,aps,showpacs,showkeys,nofootinbib]{revtex4}
\usepackage{graphicx}
\usepackage{dcolumn}
\usepackage{bm}
\usepackage{color}
\usepackage[dvipsnames]{xcolor}
\usepackage{amssymb}

\begin{document}

\title{The strong first order electroweak phase transition in the $U(1)_X$SSM}

\author{Shu-Min Zhao$^{1,2,3}$\footnote{zhaosm@hbu.edu.cn}, Jian-Fei Zhang$^{1,2,3}$\footnote{zjf09@hbu.edu.cn},
Xi-Wang$^{1,2,3}$\footnote{wangxi@stumail.hbu.edu.cn}, Xing-Xing Dong$^{1,2,3}$\footnote{dxx$\_$0304@163.com}, Tai-Fu Feng$^{1,2,3,4}$\footnote{fengtf@hbu.edu.cn}}

\affiliation{$^1$ Department of Physics, Hebei University, Baoding 071002, China}
\affiliation{$^2$ Key Laboratory of High-precision Computation and Application of Quantum Field Theory of Hebei Province, Baoding 071002, China}
\affiliation{$^3$ Research Center for Computational Physics of Hebei Province, Baoding 071002, China}
\affiliation{$^4$ Department of Physics, Chongqing University, Chongqing 401331, China}
\begin{abstract}

In the $U(1)_X$ extension of the minimal supersymmetric standard model,
there are three Higgs singlets and the corresponding trilinear terms in the Higgs effective  potential.
These new terms can allow a strongly first order electroweak phase transition(EWPT) for a wide parameter space.
We use codes CosmoTransitions to analyze the thermal evolution of the Higgs effective potential
and calculate nucleation temperature. To find reasonable
parameter spaces for strongly first order EWPT, we randomly scan many parameters, which is numerically expensive.
The diagrams are shown, that can lead to the 125 GeV Higgs mass and satisfy the first order EWPT.
This work benefits the phenomenology of $U(1)_X$SSM and exploring new physics beyond the SM.

\end{abstract}

\date{\today}

\maketitle
\section{Introduction}
Though the standard model(SM) has achieved great success for an excellent description of many experiment data
in particle physics, it still fails to explain some puzlles: 1 It can not produce tiny mass to light neutrino\cite{NeuExp};
2 It can not provide a cold dark matter candidate; 3 The observed baryon asymmetry of the universe (BAU)
is not explained in the SM\cite{EWPTSM}. On the supposition that the BAU is generated
via the electroweak baryogenesis\cite{baryogenersis1,baryogenersis2},
 the strong first order electroweak phase transition(EWPT) is necessary to provide a non-equilibrium environment\cite{chaowei,BianLG}.
 If the Higgs mass is less than 45 GeV, the strong EWPT can take place in the SM\cite{chaowei,BianLG,LiTJ}.
 However, it conflicts with the present experiment data
  for the lightest CP-even Higgs mass $m_{h^0}=125$ GeV.
 The SM CP-violation in the CKM matrix is so small that it is not able to generate a sufficient baryon asymmetry during the EWPT\cite{EWPTSM}.
   To solve this problem, the extension of SM with extra Higgs,
  heavy fermions and supersymmetric extensions of SM are possible ways\cite{LiTJ}.

During the popular models of new physics, the minimal supersymmetric extension of the standard model(MSSM)\cite{MSSM}
is a favorite one, which has been well studied for many years.
In the MSSM, there are additional sources of CP violation: the phases of $\mu$ and
supersymmetric breaking parameters. To generate a strong first order EWPT, the lightest stop quark mass should be lighter
than the top quark mass $m_t\sim 173$ GeV, that is called as the light stop scenario\cite{EWPTMSSM}. However, the current experiment
constraint for the lightest stop quark mass is $m_{\tilde{t}}>1100$ GeV\cite{2020pdg}.
So, this condition is ruled out by LHC constraints on stop masses.

With addition of the singlet $S$, the next-to-mimimal
supersymmetric standard model(NMSSM)\cite{NMSSM} has a trilinear term $\lambda A_\lambda S H_u H_d$
in the Higgs potential. In this condition, a strong enough
first order EWPT is allowed to occur\cite{EWPTnMSSM}. In U(1) gauge extensions of the MSSM (such as
UMSSM and MSSM with $U(1)^\prime$ symmetry), the EWPT is strongly
first order for a wide parameter space\cite{ref2}, whose price is introducing
new extra singlet scalars, or adding new extra heavy singlet fermions.

Taking into account the shortcomings of MSSM such as: $\mu$ problem and neutrino with zero mass,
 physicists extend MSSM and obtain many new supersymmetric models,
 where the U(1) extension is an interesting type\cite{UMSSM}. There are some works of the strong first
  order EWPT in the U(1) extensions of MSSM\cite{EWPTinU}.
  In this work, we add three Higgs singlets $\eta,~\bar{\eta},~S$ and three generation right-handed neutrinos
 to the $U(1)_X$ extension of MSSM. This model is called as $U(1)_X$SSM with the local group $SU(3)_C\otimes
SU(2)_L \otimes U(1)_Y\otimes U(1)_X$\cite{ZSMJHEPNPB,SARAH}. The right-handed neutrinos and the added Higgs singlets
produce several effects: Light neutrinos obtain tiny masses through see-saw mechanism; Right-handed neutrino
possesses dark matter character; Scalar neutrino can be dark matter candidate.
Comparing with MSSM, the so called little hierarchy problem in $U(1)_X$SSM is relived because of the added superfields.

In the superpotential of $U(1)_X$SSM, there are two terms: $\mu\hat{H}_u\hat{H}_d$ and $\lambda_H\hat{S}\hat{H}_u\hat{H}_d$.
Considering $\hat{S}$ with a non-zero VEV ($v_S/\sqrt{2}$), an effective $\mu_{eff}=\mu+\lambda_Hv_S/\sqrt{2}$ is obtained.
So, it can relieve the $\mu$ problem. The $U(1)_X$SSM has three Higgs singlets and the corresponding trilinear terms.
In the soft breaking terms, there are $B_SS^2,~L_SS,~\frac{T_\kappa}{3}S^3,
~T_{\lambda_C}S\eta\bar{\eta},~\epsilon_{ij}T_{\lambda_H}SH_d^iH_u^j$, which appear in the Higgs effective potential.
These new terms especially the trilinear terms ( $\frac{T_\kappa}{3}S^3,
~T_{\lambda_C}S\eta\bar{\eta},~\epsilon_{ij}T_{\lambda_H}SH_d^iH_u^j$ ) can
allow a strongly first-order EWPT for a wide parameter space. We use codes CosmoTransitions \cite{CT} to
analyze the thermal evolution of the effective potential
and calculate nucleation temperature.
This model has more CP-violating sources than MSSM and can generate
sufficient baryon asymmetry during EWPT.

At the critical temperature,
the role of the global minimum of the potential passes from one local minimum to another, that is
 a necessary condition for a first order phase transition. However,
 the critical temperature calculation does not account for the probability
of the first order phase transition actually taking place. Via bubble nucleation,
 first order phase transitions proceed. For the system transitioning from the
false vacuum to the true vacuum, the probability is calculated through the bounce action\cite{PA}, the Euclidean
space-time integral over the effective Lagrangian.
The authors\cite{SB} find that analyzing only the vacuum structure
via the critical temperatures can provide a misleading picture of the phase transition
patterns, and of the parameter space. So it is
important to calculate nucleation temperature to judge a successful
strong first order EWPT.

In section 2, we introduce the main content of $U(1)_X$SSM. The temperature corrections for the particle masses and
the one loop effective potential at finite temperature are given out in section 3. We study the numerical results by codes CosmoTransitions and plot
the figures in section 4. The discussion and conclusion are shown in the last section.

\section{The $U(1)_X$SSM}

Extending the local gauge group to $SU(3)_C\otimes
SU(2)_L \otimes U(1)_Y\otimes U(1)_X$ and introducing three-generation right-handed neutrinos and three Higgs singlets to MSSM,
we obtain the $U(1)_X$ extension of MSSM, which is called as $U(1)_X$SSM.
The right-handed neutrinos and Higgs singlets can solve the problem of light neutrino mass and mixing.
The CP-even parts of
the singlets $\eta,~\bar{\eta}$ and $S$ mix with the corresponding parts of $H_u$ and $H_d$.
Then the mass squared matrix of neutral CP-even Higgs is extended to $5\times5$. The introduction of $S$ can improve the lightest
CP even Higgs mass at tree level. One can find the particle contents  in our previous work\cite{ZSMJHEPNPB}.

For $U(1)_X$SSM, the superpotential reads as
\begin{eqnarray}
&&W=l_W\hat{S}+\mu\hat{H}_u\hat{H}_d+M_S\hat{S}\hat{S}-Y_d\hat{d}\hat{q}\hat{H}_d-Y_e\hat{e}\hat{l}\hat{H}_d+\lambda_H\hat{S}\hat{H}_u\hat{H}_d
\nonumber\\&&\hspace{1.0cm}+\lambda_C\hat{S}\hat{\eta}\hat{\bar{\eta}}+\frac{\kappa}{3}\hat{S}\hat{S}\hat{S}+Y_u\hat{u}\hat{q}\hat{H}_u+Y_X\hat{\nu}\hat{\bar{\eta}}\hat{\nu}
+Y_\nu\hat{\nu}\hat{l}\hat{H}_u.
\end{eqnarray}

The two Higgs doublets are same as those in MSSM, \begin{eqnarray}
&&H_{d}=\left(\begin{array}{c}H_d^0\\H_{d}^-\end{array}\right),~~~~~~~~~~~~~~~~~~~~~~H_{u}=\left(\begin{array}{c}H_{u}^+
\\H_u^0\end{array}\right),\nonumber\\&&
H_d^0={v_{d}+\phi_{d}^0+iP_{d}^0\over\sqrt{2}},~~~~~~~~~~~~~~~H_u^0={v_{u}+\phi_{u}^0+iP_{u}^0\over\sqrt{2}}.
\end{eqnarray}
$\tan\beta=v_u/v_d$ is defined by the VEVs of the Higgs superfields $H_u$ and  $H_d$.

The concrete forms of  three Higgs singlets read as
\begin{eqnarray}
&&\eta={v_{\eta}+\phi_{\eta}^0+iP_{\eta}^0\over\sqrt{2}},~~~~~~
\bar{\eta}={v_{\bar{\eta}}+\phi_{\bar{\eta}}^0+iP_{\bar{\eta}}^0\over\sqrt{2}},~~~~~~
S={v_{S}+\phi_{S}^0+iP_{S}^0\over\sqrt{2}}.
\end{eqnarray}
$v_\eta$,~ $v_{\bar\eta}$ and $v_S$ are the VEVs of the Higgs superfields $\eta$, $\bar{\eta}$ and $S$
 respectively. The $\beta_\eta$ is defined as $\tan\beta_\eta=v_{\bar{\eta}}/v_{\eta}$.

The soft SUSY breaking terms of this model are shown as
\begin{eqnarray}
&&\mathcal{L}_{soft}=\mathcal{L}_{soft}^{MSSM}-B_SS^2-L_SS-\frac{T_\kappa}{3}S^3-T_{\lambda_C}S\eta\bar{\eta}
+\epsilon_{ij}T_{\lambda_H}SH_d^iH_u^j\nonumber\\&&
-T_X^{IJ}\bar{\eta}\tilde{\nu}_R^{*I}\tilde{\nu}_R^{*J}
+\epsilon_{ij}T^{IJ}_{\nu}H_u^i\tilde{\nu}_R^{I*}\tilde{l}_j^J
-m_{\eta}^2|\eta|^2-m_{\bar{\eta}}^2|\bar{\eta}|^2\nonumber\\&&
-m_S^2S^2-(m_{\tilde{\nu}_R}^2)^{IJ}\tilde{\nu}_R^{I*}\tilde{\nu}_R^{J}
-\frac{1}{2}\Big(M_S\lambda^2_{\tilde{X}}+2M_{BB^\prime}\lambda_{\tilde{B}}\lambda_{\tilde{X}}\Big)+h.c~~.
\end{eqnarray}
We use $Y^{Y(X)}$ to denote $U(1)_{Y(X)}$ charge, and the numbers of $Y^{Y(X)}$ for the superfields are given out in our previous work\cite{ZSMJHEPNPB}.
We have proven that $U(1)_X$SSM is anomaly free. The gauge kinetic mixing is a new effect, which
 is produced by two Abelian groups $U(1)_Y$ and $U(1)_X$.

In the $U(1)_X$SSM, the covariant derivatives can be expressed as \cite{COD}
\begin{eqnarray}
&&D_\mu=\partial_\mu-i\left(\begin{array}{cc}Y^Y,&Y^X\end{array}\right)
\left(\begin{array}{cc}g_{Y},&g{'}_{{YX}}\\g{'}_{{XY}},&g{'}_{{X}}\end{array}\right)
\left(\begin{array}{c}A_{\mu}^{\prime Y} \\ A_{\mu}^{\prime X}\end{array}\right)\;.
\label{gauge1}
\end{eqnarray}
$A_{\mu}^{\prime Y}$ and $A^{\prime X}_\mu$ are the gauge fields of $U(1)_Y$ and $U(1)_X$.
Because the two Abelian gauge groups are unbroken, we can rotate the gauge coupling matrix with R
 \cite{COD} to make one non-diagonal element zero.

\begin{eqnarray}
&&\left(\begin{array}{cc}g_{Y},&g{'}_{{YX}}\\g{'}_{{XY}},&g{'}_{{X}}\end{array}\right)
R^T=\left(\begin{array}{cc}g_{1},&g_{{YX}}\\0,&g_{{X}}\end{array}\right)\;.
\label{gauge2}
\end{eqnarray}

Three gauge bosons $A^{X}_\mu,~A^Y_\mu$ and $V^3_\mu$ mix together
 and produce a $3\times3$ mass squared matrix for neutral gauge bosons\cite{g2U1X}.
 To diagonalize this matrix, two mixing angles $\theta_{W}$ and $\theta_{W}'$ are needed.
  $\sin^2\theta_{W}^\prime$ is defined as\cite{g2U1X}
\begin{eqnarray}
\sin^2\theta_{W}'=\frac{1}{2}-\frac{((g_{YX}+g_X)^2-g_{1}^2-g_{2}^2)v^2+
4g_{X}^2\xi^2}{2\sqrt{((g_{YX}+g_X)^2+g_{1}^2+g_{2}^2)^2v^4+8g_{X}^2((g_{YX}+g_X)^2-g_{1}^2-g_{2}^2)v^2\xi^2+16g_{X}^4\xi^4}}.
\end{eqnarray}

 The eigenvalues of the mass squared matrix for neutral gauge bosons are deduced. One is zero mass corresponding to  photon.
 The other two values are for $Z$ and $Z^\prime$
\begin{eqnarray}
&&m_{Z,{Z^{'}}}^2=\frac{1}{8}\Big((g_{1}^2+g_2^2+(g_{YX}+g_X)^2)v^2+4g_{X}^2\xi^2 \nonumber\\
&&\mp\sqrt{(g_{1}^2+g_{2}^2+(g_{YX}+g_X)^2)^2v^4+8((g_{YX}+g_X)^2-g_{1}^2-
g_{2}^2)g_{X}^2v^2\xi^2+16g_{X}^4\xi^4}\Big).
\end{eqnarray}
Here, $v=\sqrt{v_u^2+v_d^2}$ and $\xi=\sqrt{v_{\eta}^2+v_{\bar{\eta}}^2}$.

  At tree level, the Higgs potential is deduced\cite{ZSMJHEPNPB}
\begin{eqnarray}
&&V_{0}=\frac{1}{2}g_X(g_X+g_{YX})(|H_d^0|^2-|H_u^0|^2)(|\eta|^2-|\bar{\eta}|^2)+|\lambda_H|^2|H_u^0H_d^0|^2+m^2_{s}|S|^2\nonumber\\&&
+\frac{1}{8}\Big(g_1^2+g_2^2+(g_X+g_{YX})^2\Big)(|H_d^0|^2-|H_u^0|^2)^2+\frac{1}{2}g_X^2(|\eta|^2-|\bar{\eta}|^2)^2+\lambda_C^2|\eta\bar{\eta}|^2
\nonumber\\&&
+(|\mu|^2+|\lambda_H|^2|S|^2+2\mathrm{Re}[\mu^*\lambda_HS])(|H_d^0|^2+|H_u^0|^2)+|\lambda_C|^2|S|^2(|\eta|^2+|\bar{\eta}|^2)
\nonumber\\&&+2\mathrm{Re}[l_W^*(2M_SS+\lambda_C\eta\bar{\eta}-\lambda_HH_u^0H_d^0
+\kappa S^2)]+4|M_S|^2|S|^2+
2\mathrm{Re}[\lambda_C^*\kappa\eta^*\bar{\eta}^*S^2]
\nonumber\\&&+|\kappa|^2|S|^4+4\mathrm{Re}[M_S^*S^*(\lambda_C\eta\bar{\eta}-\lambda_HH_u^0H_d^0+\kappa S^2)]
-2\mathrm{Re}[\lambda_C^*\lambda_H\eta^*\bar{\eta}^*H_u^0H_d^0]+|l_W|^2\nonumber\\&&-2\mathrm{Re}[B_\mu H_d^0H_u^0]+ 2\mathrm{Re}[L_S S]
+\frac{2}{3}\mathrm{Re}[T_kS^3]+ 2\mathrm{Re}[T_{\lambda_C}\eta\bar{\eta}S]-2\mathrm{Re}[T_{\lambda_H}H_d^0H_u^0 S]\nonumber\\&&-2\mathrm{Re}[\lambda_H\kappa^* H_u^0H_d^0 (S^2)^*]
+m^2_\eta|\eta|^2+m^2_{\bar{\eta}}|\bar{\eta}|^2+
m^2_{H_u^0}|H_u|^2+m^2_{H_d}|H_d|^2+2\mathrm{Re}[B_S S^2].\label{Vtree}
\end{eqnarray}

 The parameters
($\mu,~\lambda_H,~\lambda_C,~l_W,~ M_S,~B_\mu,~L_S,~T_{\kappa},~ T_{\lambda_C},~ T_{\lambda_H},~\kappa,~B_S$) in Eq.
(\ref{Vtree}) are supposed as real parameters to simplify the discussion. Through the formula
\begin{eqnarray}
&&\left \langle \frac{\partial V_{tree}}{\partial H_u^0} \right \rangle=\left \langle \frac{\partial V_{tree}}{\partial H_d^0} \right \rangle=
\left \langle \frac{\partial V_{tree}}{\partial \eta} \right \rangle=\left \langle \frac{\partial V_{tree}}{\partial \bar{\eta}} \right \rangle=\left \langle \frac{\partial V_{tree}}{\partial S} \right \rangle=0,
\end{eqnarray}
 one can obtain the following tadpole equations\cite{ZSMJHEPNPB}
\begin{eqnarray}
&&\frac{v_d^2-v_u^2}{8}\Big(g_1^2+g_2^2+(g_X+g_{YX})^2\Big)+
\frac{g_X}{4}(g_X+g_{YX})(v^2_{\eta}-v^2_{\bar{\eta}})+
\mu^2+\frac{\lambda_H^2}{2}(v_u^2+v_S^2)+m_{H_d}^2\nonumber\\&&+\sqrt{2}\mu\lambda_H v_S
-[\lambda_H(\sqrt{2}M_Sv_S+l_W+\frac{\lambda_C}{2}v_{\eta}v_{\bar{\eta}}
+\frac{\kappa}{2} v_S^2)+B_\mu +\frac{T_{\lambda_H}}{\sqrt{2}}v_S]\tan\beta=0,\\&&
\frac{v_u^2-v_d^2}{8}\Big(g_1^2+g_2^2+(g_X+g_{YX})^2\Big)+
\frac{g_X}{4}(g_X+g_{YX})(v^2_{\bar{\eta}}-v^2_{\eta})+
\mu^2+\frac{\lambda_H^2}{2}(v_S^2+v_d^2)+m_{H_u}^2\nonumber\\&&+\sqrt{2}\mu\lambda_H v_S
-[\lambda_H(\sqrt{2}M_Sv_S+l_W+\frac{\lambda_C}{2}v_{\eta}v_{\bar{\eta}}
+\frac{\kappa}{2} v_S^2)+B_\mu +\frac{T_{\lambda_H}}{\sqrt{2}}v_S]\cot\beta=0,
\\&&\frac{g_X^2}{2}(v_{\eta}^2-v_{\bar{\eta}}^2)-
\frac{g_X}{4}(g_X+g_{YX})(v^2_{u}-v^2_{d})
+\frac{\lambda_C^2}{2}(v_S^2+v_{\bar{\eta}}^2)+m_{\eta}^2
+[\lambda_C(l_W+\sqrt{2}M_Sv_S\nonumber\\&&-\frac{1}{2}\lambda_Hv_{u}v_d
+\frac{1}{2}\kappa  v_S^2)+\frac{T_{\lambda_C}}{\sqrt{2}}v_S]\tan\beta_\eta=0,
\\&&
\frac{1}{2}g_X^2(v_{\bar{\eta}}^2-v_{\eta}^2)+
\frac{1}{4}g_X(g_X+g_{YX})(v^2_{u}-v^2_{d})
+\frac{1}{2}\lambda_C^2 (v_S^2+v_\eta^2)+m_{\bar{\eta}}^2
+[\lambda_C(l_W+\sqrt{2}M_Sv_S\nonumber\\&&-\frac{1}{2}\lambda_Hv_{u}v_d+\frac{1}{2}\kappa  v_S^2)+\frac{T_{\lambda_C}}{\sqrt{2}}v_S]\cot\beta_\eta=0,
\\&&\frac{\lambda_H^2}{2}v^2+\frac{\lambda_C^2}{2}\xi^2
+4M_S^2+\kappa(\kappa v_S^2 +2l_W+3\sqrt{2}M_S v_S+\lambda_C v_\eta v_{\bar{\eta}}-\lambda_H v_u v_d)
+\frac{T_k v_S}{\sqrt{2}}+ 2B_S\nonumber\\&&+m_S^2+[L_{S}
+M_S(2l_W+\lambda_Cv_\eta v_{\bar\eta}-\lambda_H v_u v_d)
+\frac{\mu\lambda_Hv^2+T_{\lambda_C} v_{\eta}v_{\bar{\eta}}-T_{\lambda_H} v_u v_d}{2}]\frac{\sqrt{2}}{v_S}=0.
\end{eqnarray}

\section{The one loop effective potential at finite temperature}
To simplify the discussion, we change the tree level potential $V_0$ in Eq.(\ref{Vtree}) to the form $V_{0}(h,y,z)$
with the relations
\begin{eqnarray}
&&h^2=(\phi_d^0)^2+(\phi_u^0)^2,~~~~~~y^2=(\phi_\eta^0)^2+(\phi_{\bar{\eta}}^0)^2,
\nonumber\\&&z^2=(\phi^0_S)^2, ~~~~\frac{\phi_u^0}{\phi_d^0}=\tan\beta, ~~~~\frac{\phi_{\bar{\eta}}^0}{\phi_\eta^0}=\tan\beta_\eta.
\end{eqnarray}
The one loop effective potential at finite temperature\cite{SMVT} can be written in the following form\cite{LiTJ}
\begin{eqnarray}
V_{eff}(h,y,z,T)=V_{0}(h,y,z)+V_{1}(h,y,z,0)+\Delta V_{1}(h,y,z,T)+\Delta V_{daisy}(h,y,z,T)\label{oneeff}.
\end{eqnarray}
Here, $V_{0}(h,y,z)$ is the tree level potential.
The one loop zero temperature correction is represented by $V_{1}(h,y,z,0)$\cite{oneloopT0V}.
$\Delta V_{1}(h,y,z,T)$ represents the temperature dependent one loop correction\cite{loopT}, while $\Delta V_{daisy}(h,y,z,T)$ denotes the multi-loop daisy correction\cite{BosonRing}.

The concrete forms of $V_{1}(h,y,z,0),~\Delta V_{1}(h,y,z,T)$ and $\Delta V_{daisy}(h,y,z,T)$ are shown explicitly
\begin{eqnarray}
&&V_{1}(h,y,z,0)=\sum_i\frac{n_i}{64\pi^2}m_i^4(h,y,z)\Big(
\log\frac{m_i^2(h,y,z)}{Q^2}-C_i\Big),\nonumber\\&&
\Delta V_{1}(h,y,z,T)=\frac{T^4}{2\pi^2}\Big\{
\sum_in_iJ_i\Big[\frac{m_i^2(h,y,z)}{T^2}\Big]\Big\},\nonumber\\&&
\Delta V_{daisy}(h,y,z,T)=-\frac{T}{12\pi}\sum_{i=bosons}n_i[\mathcal{M}_i^3(h,y,z,T)-m_i^3(h,y,z)].\label{VVT}
\end{eqnarray}
In the zero temperature correction, $Q$ is the renormalization scale and supposed at TeV order.
 $m_i(h,y,z)$ denote field-dependent masses and $n_i$ are the number of degrees of freedom.
In Eq.(\ref{VVT}), the particle masses $m_i$  include Fermions and Bosons.
The considered Fermions are quarks($t,~b$), lepton($\tau$),
charginos, neutralinos and neutrinos. While, the considered Bosons are
 up-type squarks,  down-type squarks, sleptons,
CP-even sneutrinos, CP-odd sneutrinos, CP-even Higgs, CP-odd Higgs, Goldstones, vector Bosons($W^\pm,~Z,~Z^\prime$).
$n_i$ are the degrees of freedom for the corresponding mass eigenstates.
In $U(1)_X$SSM, the concrete values for $n_i$ are the following:
for quarks $n_i=-12$, for leptons and charginos $n_i=-4$, for neutralinos and neutrinos $n_i=-2$;  for
squarks $n_i=6$, for sleptons and charged Higgs(Goldstones) $n_i=2$, for $Z$ and $Z^\prime$ Bosons $n_i=3$,
for $W$ Boson $n_i=6$. for CP-odd, CP-even sneutrinos and the remaining Higgs scalars(Goldstones) $n_i=1$.
The contents $C_i$ depend on the regularization scheme. In the $\overline{MS}$ scheme, they are assumed as $C_i=\frac{3}{2}$ for scalars and fermions and $C_i=\frac{5}{6}$ for gauge bosons. There is no evidence of the Goldstone catastrophe in the potential of this model.
As discussed in Refs.\cite{rs1,rs2,rs3}, the IR divergences are spurious and can be tamed through resummation.

For bosons and fermions, the $J_i$ functions in the one loop effective potential at finite temperature have different forms\cite{LiTJ,TRPT}
\begin{eqnarray}
&&J_B\Big[m_B^2(h,y,z)/T^2\Big]=\int_0^\infty dx ~ x^2\log\Big\{1-\exp\Big[-\sqrt{x^2+m_B^2(h,y,z)/T^2}\Big]\Big\},
\nonumber\\&&
J_F\Big[m_F^2(h,y,z)/T^2\Big]=\int_0^\infty dx ~ x^2\log\Big\{1+\exp\Big[-\sqrt{x^2+m_F^2(h,y,z)/T^2}\Big]\Big\}.
\end{eqnarray}
At high temperature and low temperature, the functions $J_B[m_B^2(h,y,z)/T^2]$ and $J_F[m_F^2(h,y,z)/T^2]$ can
be expanded\cite{chaowei,BianLG,LiTJ}. In the numerical calculation of Ref.\cite{LiTJ}, the authors give perfect approximations for the functions $J_B[m_B^2(h,y,z)/T^2]$ and $J_F[m_F^2(h,y,z)/T^2]$.

Adding temperature dependent self-energy contributions $\Pi(T)$ to $m_i^2(h,y,z)$, one can obtain the temperature dependent scalar mass squared $\mathcal{M}^2(h,y,z,T)=m^2(h,y,z)+\Pi(T)$\cite{BTmass}. In this equation, $\Pi(T)$ is proportional to $T^2$. The longitudinal components of gauge bosons receive such contributions. The $\Pi(T)$ for particles in $U(1)_X$SSM are shown here.
Following the method\cite{LiTJ,BTmass} for the temperature correction of particle mass,
 we deduce Eqs.(20$\sim$23) in our model.

1. $\Pi(T)$ for scalar quarks
\begin{eqnarray}
&&\Pi_{\tilde{Q}_i}(T)=\Big(\frac{2}{3}g_3^2+\frac{3}{8}g_2^2+\frac{1}{72}g_1^2
+\frac{1}{4}(Y_{u_i}^2+Y_{d_i}^2)\Big)T^2,\nonumber\\&&
\Pi_{\tilde{u}_R^i}(T)=\Big(\frac{2}{3}g_3^2+\frac{2}{9}g_1^2
+\frac{1}{2}Y_{u_i}^2+\frac{1}{8}g_X^2\Big)T^2,\nonumber\\&&
\Pi_{\tilde{d}^i_R}(T)=\Big(\frac{2}{3}g_3^2+\frac{1}{18}g_1^2
+\frac{1}{2}Y_{d_i}^2+\frac{1}{8}g_X^2\Big)T^2.
\end{eqnarray}
2. $\Pi(T)$ for scalar charged leptons and scalar neutrinos
\begin{eqnarray}
&&\Pi_{\tilde{L}_i}(T)=\Big(\frac{3}{8}g_2^2+\frac{1}{8}g_1^2
+\frac{1}{4}Y_{e_i}^2\Big)T^2,~~
\Pi_{\tilde{e}_R^i}(T)=\Big(\frac{1}{2}g_1^2
+\frac{1}{2}Y_{e_i}^2+\frac{1}{8}g_X^2\Big)T^2,\nonumber\\&&
\Pi_{\tilde{\nu}_R^i}(T)=\Big(\frac{1}{4}Y_{X}^2+\frac{1}{8}g_X^2\Big)T^2.
\end{eqnarray}
3. $\Pi(T)$ for Higgs doublets and singlets
\begin{eqnarray}
&&\Pi_{H_d}(T)=\Big(\frac{3}{8}g_{2}^2+\frac{1}{8}g_1^2+\frac{3}{4}Y_b^2+\frac{1}{8}g_X^2
+\frac{1}{4}Y^2_{e_3}+\frac{1}{4}\lambda_H^2\Big)T^2,\nonumber\\&&
\Pi_{H_u}(T)=\Big(\frac{3}{8}g_{2}^2+\frac{1}{8}g_1^2+\frac{3}{4}Y_t^2+\frac{1}{8}g_X^2
+\frac{1}{4}\lambda_H^2\Big)T^2,~~~
\Pi_{\eta}(T)=\Big(\frac{1}{2}g_{X}^2+\frac{1}{4}\lambda_C^2\Big)T^2,\nonumber\\&&
\Pi_{\bar{\eta}}(T)=\Big(\frac{1}{2}g_{X}^2+\frac{1}{4}\lambda_C^2+\frac{1}{4}\lambda_X^2\Big)T^2,~~~~~~
\Pi_{S}(T)=\Big(\frac{1}{4}\kappa^2+\frac{1}{4}\lambda_C^2+\frac{1}{2}\lambda_H^2\Big)T^2.
\end{eqnarray}
4.  $\Pi(T)$ for the longitudinal components of gauge bosons
\begin{eqnarray}
&&\Pi_{g_3}(T)=\frac{9}{2}g_3^2T^2,~~~~~~~~~~~~\Pi_{g_2}(T)=\frac{9}{2}g_2^2T^2,\nonumber\\&&
\Pi_{g_1}(T)=\frac{11}{2}g_1^2T^2,~~~~~~~~~~\Pi_{g_X}(T)=\frac{9}{2}g_X^2T^2.
\end{eqnarray}

At finite temperature, the effective potential receives the thermal corrections.
The tree level cubic term and the loop corrections can produce interesting effects on the phase transition.
To study the first-order EWPT better, we do not adopt the high temperate approximation.
Using the codes CosmoTransitions, we research the one loop effective potential at finite
temperature shown as Eq.(\ref{oneeff}).
The codes CosmoTransitions can calculate the important parameters of phase transition,
such as the critical temperature, the nucleation temperature, the step and type of phase transition, and the action, etc.

The phase transition can be a
first order EWPT, because there is a potential with barrier between the two minima. It is a tunneling process.
Through nucleations of electroweak bubbles which expand, collide and coalesce, the transition proceeds and in the end the universe turns into electroweak symmetry breaking phase. Through the first order EWPT, baryon asymmetry can be generated from electroweak baryogenesis. The sphaleron process in the bubble should be sufficiently suppressed so as to preserve the generated baryon asymmetry after the EWPT.
This requirement can be expressed as \cite{LiTJ}
\begin{eqnarray}
 \frac{v(T_n)}{T_n}\gtrsim 1,
\end{eqnarray}
where $T_n$ is the nucleation temperature. In the SM, $v(T)$ represents the VEV of the Higgs field $H^0$ at
temperature T.
In the MSSM, the condition is similar
\begin{eqnarray}
v(T)=\sqrt{v_d^2(T)+v_u^2(T)}.
\end{eqnarray}
Here, $v_d(T)$ and $v_u(T)$ are the VEVs of the two neutral Higgs $H^0_d$ and $H^0_u$.
The condition of $U(1)_X$SSM is more complex than that of MSSM, because three Higgs singlets are added.

Bubble nucleation is a random event. There is always a possibility of bubble nucleation.
 If the following condition is satisfied, there is one bubble generate in a Hubble volume, and the transition is able to complete.
\begin{eqnarray}
S_E(T_n)/T_n\simeq140,
\end{eqnarray}
$S_E$ represents the Euclidean bubble action.

\section{numerical results}
Considering our previous works in the $U(1)_X$SSM\cite{ZSMJHEPNPB}, we study the numerical results in this section.
The mass of the new neutral gauge boson is strict, and we take $M_{Z^\prime}>4.5~{\rm TeV}$\cite{ZSMJHEPNPB}.
It is at $99\%$ CL, that $\frac{M_{Z^\prime}}{g_X}\geq 6{\rm TeV}$\cite{6TeV} for the ratio between $M_{Z^\prime}$
and its gauge coupling. LHC experiment gives constraint for the new angle $\beta_\eta$ as $\tan\beta_\eta<1.5$\cite{betaeta}.
As a concrete example, we take the following parameters as
\begin{eqnarray}
&& \tan\beta=20,~
Y_X=1,~M_1=0.8~{\rm TeV},~M_2=1.2~{\rm TeV},~T_{\kappa}=1.6~{\rm GeV},\nonumber\\&&
M_{BL}=T_{L_{ii}}=T_{\nu_{ii}}=T_X=T_{u_{ii}}=T_{d_{ii}}=1~{\rm TeV},~
B_S = -1~ {\rm TeV}^2,\nonumber\\&&B_\mu = 1~{\rm TeV}^2,~
M_{\tilde{Q}_{ii}}^2=3.5~{\rm TeV}^2,~M_{\tilde{U}_{ii}}^2=3~{\rm TeV}^2,~M_{\tilde{D}_{ii}}^2=4.5~{\rm TeV}^2,\nonumber\\&&M_{\tilde{L}_{ii}}^2=0.5~{\rm TeV}^2,
~ M_{\tilde{E}_{ii}}^2=3~{\rm TeV}^2,~ M_{\tilde{\nu}_{ii}}^2=0.2~{\rm TeV}^2,~l_W = 4~{\rm TeV}^2.
\label{parameter}
\end{eqnarray}
In order to find the parameter space satisfying first order EWPT, we use the following parameters as variables with their value ranges.
\begin{eqnarray}
&&0.3\leq g_X\leq0.8,~0.01\leq g_{YX}\leq0.5,~0.1\leq\kappa\leq1.1,~-3~{\rm TeV}\leq T_{\lambda_C}\leq3~{\rm TeV},
\nonumber\\&&-3~{\rm TeV}\leq T_{\lambda_H}\leq3~{\rm TeV},~0.5~{\rm TeV}\leq M_S\leq 4~{\rm TeV},~
-1\leq\lambda_H\leq1,~-1\leq\lambda_C\leq1,\nonumber\\&&0.6~{\rm TeV}\leq\mu\leq1.5~{\rm TeV} ,~-10~{\rm TeV}^3\leq L_{S}\leq10
~{\rm TeV}^3,~0.75 \leq\tan\beta_{\eta}\leq 0.99.
\end{eqnarray}
Some parameters such as $m_{H_u}^2,~m_{H_d}^2,~m^2_{\eta},~m^2_{\bar{\eta}}$ can be calculated from the tadpole equations\cite{ZSMJHEPNPB}
and the zero temperature correction.
Using the codes CosmoTransitions, we study the phase transition in the $U(1)_X$SSM.
We do not collect the results of the second phase transition, because they are not useful for
 the first order EWPT and the numerical results are so expensive for working time.
 The showed plots are all suit for the first order EWPT.

\begin{table}
\caption{ The markers in numerical results}\label{TAB}
\begin{tabular}{|c|c|c|c|}
\hline
Shape style & weak first order EWPT & strong first order EWPT & 125 GeV Higgs mass\\
\hline
\textcolor{red} {$\bullet$} &  $\surd$  & $\times$ & $\times$ \\
\hline
\textcolor{yellow}{$\bullet$} & $\times$ & $\surd$ & $\times$  \\
\hline
\textcolor{green} {$\bullet$} &$\surd$ & $\times$ & $\surd$ \\
\hline
\textcolor{blue} {$\bullet$} & $\times$ & $\surd$ & $\surd$ \\
\hline
\end{tabular}
\label{quarks}
\end{table}

\begin{figure}[h]
\setlength{\unitlength}{5.0mm}
\centering
\includegraphics[width=2.6in]{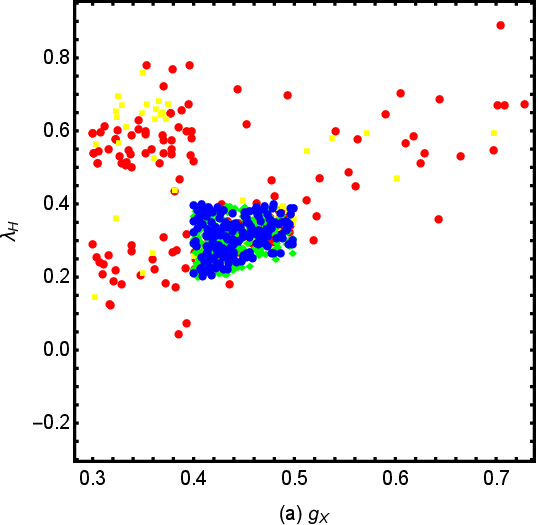}~~~~\includegraphics[width=2.6in]{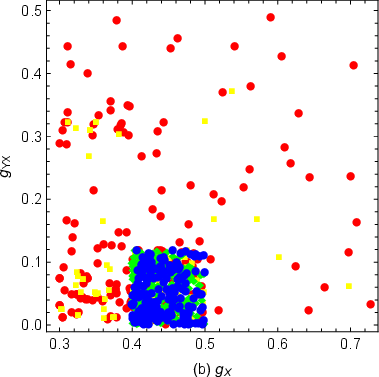}\\
\vspace{0.3cm}
\includegraphics[width=2.6in]{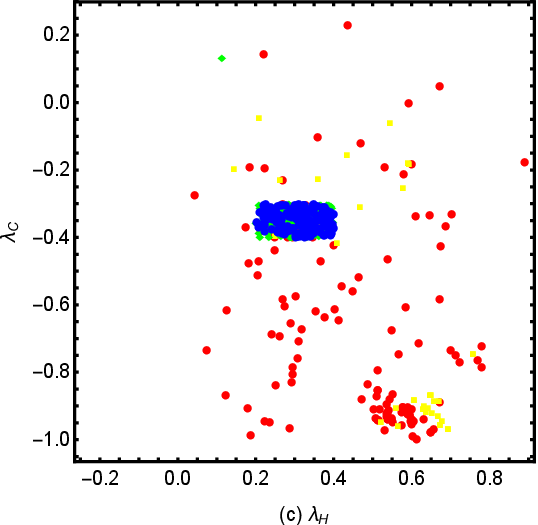}
\caption{ The diagram (a) shows the points in the plane of $g_X$ versus $\lambda_H$.
The  diagram (b) shows the points in the plane of $g_X$ versus $g_{YX}$.
The  diagram (c) shows the points in the plane of $\lambda_H$ versus $\lambda_C$.}\label{sibiaotu}
\end{figure}

In the Fig.\ref{sibiaotu}, we show the plots calculated from the codes CosmoTransitions. The meanings of \textcolor{red} {$\bullet$}, \textcolor{yellow} {$\bullet$}, \textcolor{green} {$\bullet$} and \textcolor{blue} {$\bullet$}
are collected in the table \ref{TAB}. The points plotted by \textcolor{green} {$\bullet$} and \textcolor{blue} {$\bullet$}
can lead to 125GeV Higgs mass, where \textcolor{blue} {$\bullet$} denote the parameters for strong first order EWPT
and \textcolor{green} {$\bullet$} represent the condition of the weak first order EWPT. For \textcolor{green} {$\bullet$} and \textcolor{blue} {$\bullet$}, the corresponding phase transitions are all 1 step in our obtained parameter space.
\textcolor{red} {$\bullet$} and \textcolor{yellow} {$\bullet$}
represent the points for weak and strong first order EWPT respectively,
without satisfying the constraint from 125 GeV Higgs mass.
The phase transitions denoted by \textcolor{red} {$\bullet$} and \textcolor{yellow} {$\bullet$}
include 1 step, 2 step and 3 step first order EWPT\cite{twostep}, where 1 step  first order EWPTs are dominant.
Because they do not satisfy the Higgs mass constraint, we do not further distinguish between them.
The strongly first order EWPT is of interest, and the nucleation temperature
is obtained from the codes CosmoTransitions.

The Fig.\ref{sibiaotu}(a)
shows the plots in the plane of $g_X$ and $\lambda_H$. $g_X$ is the $U(1)_X$ gauge coupling constant.
$\lambda_H$ is the constant for the term $\lambda_H\hat{S}\hat{H}_u\hat{H}_d$ in the super potential.
$g_X$ and $\lambda_H$ both appear in the mass squared matrix of CP-even Higgs at tree level.
Therefore, they are both important parameters. The four type points are in the region $\lambda_H>0$.
During $g_X$ range $0.3\leq g_X \leq 0.5$, there are more points.
\textcolor{green} {$\bullet$} and \textcolor{blue} {$\bullet$}
are concentrated in a small area with $0.4\leq g_X \leq 0.5$ and $0.2\leq \lambda_H\leq0.4$,
because these results are constrained by the 125GeV Higgs mass.
The blue area looks like a trapezium, which is better than the other area.

The Fig.\ref{sibiaotu}(b) is shown in the plane of $g_X$ and $g_{YX}$.
$g_{YX}$ is the coupling constant for gauge mixing of $U(1)_Y$ and $U(1)_{X}$, which is a new parameter beyond
MSSM and can bring new effect. Though the points appear in the almost whole region of the plane,
they concentrate in the bottom left corner with $0.3\leq g_X \leq 0.5$ and $0\leq g_{YX}\leq 0.18$.
In the square area $0.4\leq g_X \leq 0.5$ and $0\leq g_{YX}\leq 0.1$,
there are a lot of \textcolor{blue} {$\bullet$}.

In Fig.\ref{sibiaotu}(c), the four types plots are shown in the plane of $\lambda_H$ and $\lambda_C$.
$\lambda_C$ emerges in the term $\lambda_C\hat{S}\hat{\eta}\hat{\bar{\eta}}$ of the superpotential.
Because $\eta$ and $\bar{\eta}$ are Higgs singlets, the term including $\lambda_C$ give contributions to
the CP-even Higgs mass squared matrix.
Then $\lambda_C$ should influence Higgs mass to some extent.
All the points of the numerical results are scattered in most areas.
Most \textcolor{red} {$\bullet$} appear at area $0.1\leq\lambda_H\leq0.8$ and $-1.0\leq\lambda_C\leq-0.3$.
Obviously, \textcolor{blue} {$\bullet$} are concentrated in much smaller region
$0.2\leq\lambda_H\leq0.4$ and $-0.3\leq\lambda_C\leq-0.4$, because \textcolor{blue} {$\bullet$} obey
the constraint from 125GeV Higgs mass, and it is reasonable.

\begin{figure}[h]
\setlength{\unitlength}{5.0mm}
\centering
\includegraphics[width=2.6in]{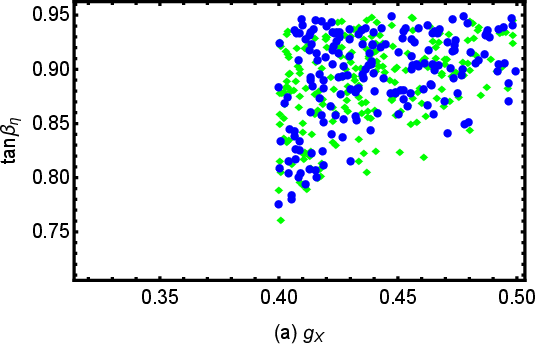}~~~~\includegraphics[width=2.6in]{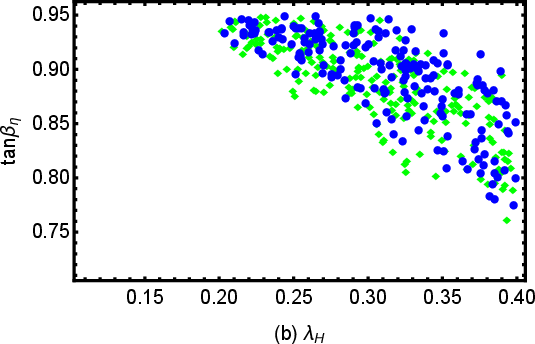}\\
\vspace{0.3cm}
\includegraphics[width=2.6in]{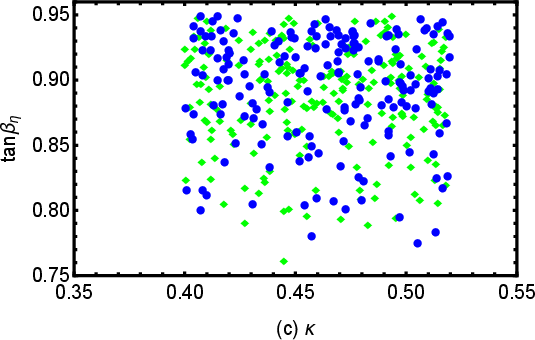}
\caption{The diagram (a) shows the points in the plane of $g_X$ versus $\tan\beta_\eta$.
The  diagram (b) shows the points in the plane of $\lambda_H$ versus $\tan\beta_\eta$.
The  diagram (c) shows the points in the plane of $\kappa$ versus $\tan\beta_\eta$.}\label{TBN}
\end{figure}

Since the results represented by
\textcolor{red} {$\bullet$} and \textcolor{yellow} {$\bullet$} do not lead to 125GeV Higgs mass,
we do not show them in the following figure.
Then, only \textcolor{blue} {$\bullet$} and
\textcolor{green} {$\bullet$} are plotted in latter analysis.
The parameter $\tan\beta_\eta$ affects the masses of many particles including Higgs,
 and appears in the Higgs potential at tree level.
 Therefore, it should bring obvious effect on the phase transition.
 The Fig.\ref{TBN}(a) embodies
 \textcolor{blue} {$\bullet$} and \textcolor{green} {$\bullet$}
 in the plane of $g_X$ versus $\tan\beta_\eta$.
 The points look like a right trapezoid in the whole.
 Most \textcolor{blue} {$\bullet$} concentrate in the
 area $0.4\leq g_X\leq0.5$ and $0.85\leq\tan\beta_\eta\leq0.95$.
 When $\tan\beta_\eta<0.85$, the both type points decrease quickly.

To see the effects of $\lambda_H$ and $\tan\beta_\eta$,
\textcolor{blue} {$\bullet$} and \textcolor{green} {$\bullet$} are
shown in  the Fig.\ref{TBN}(b).
All the points appear at the top right corner of this diagram, and most of the space is blank,
especially the bottom left corner.
It implies that the constraints from both 125GeV Higgs and first order EWPT are strict.

The  Fig.\ref{TBN}(c) is shown in the plane of $\kappa$ versus $\tan\beta_\eta$.
$\kappa$ is the parameter in the term $\frac{1}{3}\kappa\hat{S}\hat{S}\hat{S}$ of the superpotential.
$\kappa$ has relation with the Higgs
tree level potential and Higgs mass matrix through the mixing with Higgs singlet $\hat{S}$.
From analytical analysis, $\kappa$ should affect the Higgs and phase transition to some extent but not strong.
This diagram is exactly what is reflected. The dots become fewer and fewer from top to bottom.
It implies that the effect of $\tan\beta_\eta$ is stronger than that of $\kappa$ obviously.

\textcolor{blue} {$\bullet$} and \textcolor{green} {$\bullet$} represent the first order EWPT that can take place.
For these points, the nucleation temperature $T_n$ and the Euclidean bubble
action $S_E(T_n)$ are calculated through the codes CosmoTransitions.
For the weak first order EWPT points \textcolor{green} {$\bullet$},
the nucleation temperature $T_n$ is relatively high and in the region from 650 GeV to 1000 GeV.
On the other hand, the nucleation temperature $T_n$  of the
strong first order EWPT points \textcolor{blue} {$\bullet$} is more reasonable.
Most \textcolor{blue} {$\bullet$} are located in the region $100 ~{\rm GeV}\leq T_n\leq600~{\rm GeV}$,
which implies strong first order EWPT can be realized in fact.
As $T_n$ is defined as $S_E(T_n)/T_n=140$, the numerical results for the ratios $S_E(T_n)/T_n$ of all the points
\textcolor{blue} {$\bullet$} and \textcolor{green} {$\bullet$} are very close to 140.
The distribution discrepancy is due to numerical errors.

\section{discussion and conclusion}

In the $U(1)_X$ extension of MSSM, we study the strong first order EWPT. The Higgs
singlets $\hat{\eta},~\hat{\bar{\eta}}$ and $\hat{S}$ are beyond MSSM, and they bring new terms to the Higgs potential.
The one loop effective potential at finite temperature is composed of four parts: the tree level potential $V_{0}(h,y,z)$,
the one loop zero temperature correction $V_{1}(h,y,z,0)$, the temperature dependent one loop correction $\Delta V_{1}(h,y,z,T)$ and the multi-loop daisy correction $\Delta V_{daisy}(h,y,z,T)$. The tree level potential has $T_{\lambda_C}S\eta\bar{\eta},~\epsilon_{ij}T_{\lambda_H}SH_d^iH_u^j$ etc. coming from the soft breaking terms.
These terms go beyond the MSSM and allow the strong first order EWPT to take place.

In the numerical calculation, we take the parameters considering the experiment constraints
especially from the 125GeV Higgs mass.
At very high temperature, the global minimum is at the origin.
As the temperature drops down, the phase transition takes place.
Taking several parameters as variable,
we scan the parameter space that can lead to 125GeV
Higgs mass and strong first order EWPT.
1 step phase transitions are dominant,
and the nucleation temperature $T_n$ is
reasonable for the strong first order EWPT.
The effects of added Higgs singlets for phase
 transition need more work, and we shall study them in the future.

\begin{acknowledgments}
This work is supported by National Natural Science Foundation of China (NNSFC)
(No. 12075074), Natural Science Foundation of Hebei Province
(A202201022, A2022201017), Natural Science Foundation of Hebei Education Department(QN2022173).
\end{acknowledgments}

\end{document}